\documentclass[a4paper,reqno,12pt,final]{amsart}
\usepackage{amssymb,euscript,bbold}
\usepackage{array}
\usepackage[notref,notcite,color]{showkeys}
\usepackage{graphicx}
\usepackage{subfigure}
\renewcommand{\d}{\partial}
\newcommand{\half}{\tfrac12}
\newcommand{\reg}{\text{reg}}
\renewcommand{\gg}{\mathfrak g}

\newcommand{\sG}{\mathsf G}
\newcommand{\sT}{\mathsf T}

\newcommand{\1}{\mathbb 1}
\newcommand{\C}{\mathbb C}
\newcommand{\R}{\mathbb R}
\newcommand{\Z}{\mathbb Z}
\newcommand{\su}{\mathfrak{su}}
\renewcommand{\sl}{\mathfrak{sl}}
\newcommand{\N}{\EuScript N}
\newcommand{\eC}{\EuScript{C}}

\newcommand{\AdS}{\text{AdS}}
\newcommand{\SL}{\mathrm{SL}}
\newcommand{\GL}{\mathrm{GL}}

\newcommand{\SU}{\mathrm{SU}}
\newcommand{\SO}{\mathrm{SO}}
\newcommand{\eO}{\EuScript{O}}
\newcommand{\eF}{\EuScript{F}}

\DeclareMathOperator{\diag}{diag}
\DeclareMathOperator{\Ad}{Ad}
\DeclareMathOperator{\Tr}{Tr}

\newcommand{\Mat}{\mathrm{Mat}}
\begin{document}

\title[]{D-branes in an $\AdS_3$ background${}^*$} 
\author[]{Sonia Stanciu}
\address[]{\begin{flushright}Theoretical Physics Group\\
Blackett Laboratory\\
Imperial College\\
London SW7 2BZ, UK\\
\end{flushright}}
\email{s.stanciu@ic.ac.uk}
\thanks{${}^*$ Imperial/TP/98-99/36, {\tt hep-th/9901122}}
\date{\today}
\begin{abstract}
  We study the possible D-brane configurations in an $\AdS_3\times S^3
  \times T^4$ background with a NS-NS B field.  We use its WZW model
  description and the boundary state formalism, and we analyse the
  bosonic and the $N{=}1$ supersymmetric cases separately.  We also
  discuss the corresponding classical open string sigma model.  We
  determine the spacetime supersymmetry preserved by the
  supersymmetric D-brane configurations.
\end{abstract}
\maketitle

\section{Introduction}

In the framework of the AdS/SYM correspondence conjectured by
Maldacena in \cite{M}, important particles and vacuum configurations
(including interaction vertices) in the gauge theory are described, on
the string theory side, by D-brane configurations (wrapped or not) in
the corresponding type IIB background.  Thus in \cite{Wbaryon} Witten
showed how various D-brane configurations in the $\AdS_5\times S^5$
and $\AdS_5\times \R P^5$ backgrounds are associated with the baryon
vertex and the Pfaffian particle, respectively, in the
four-dimensional gauge theory at the boundary of $\AdS_5$.  Later, in
\cite{Im,CGS}, such D-brane configurations were obtained as classical
solutions of the Born-Infeld action.

Although the general belief seems to be that the AdS/SYM connection
only makes sense in the Higgs branch of the SYM theory, a proposal was
made \cite{DT} to identify certain vacuum configurations of the large
$N$ limit of the $\N{=}4$ $\SU(N)$ SYM corresponding to the Coulomb
branch with particular D3-brane configurations in the bulk of the
$\AdS_5\times S^5$, whereby the D3-branes have worldvolumes parallel
to the AdS boundary.  This suggestion was further pursued in \cite{BC}
where, by using a low energy analysis, more general D3-brane
configurations were analysed in the AdS bulk, and the corresponding
amount of supersymmetry preserved by these D-brane configurations was
determined.

In the face of these facts it appears desirable to carry out a
microscopic SCFT analysis of the possible D-brane configurations in
various AdS-type backgrounds.
Such an analysis however would rely essentially on the explicit
knowledge of the CFT underlying these string backgrounds.
Unfortunately, despite the progress made in constructing type IIB
string theories with RR background fields on AdS spaces
\cite{MT1,P5,KR5,P3,RR,PR} (in the framework of the Green-Schwarz
formalism), we are still lacking a satisfactory worldsheet description
for most of these backgrounds.  The only tractable case so far remains
that of type IIB string theory on $\AdS_3\times K$, where $K$ is a
compact manifold, and we allow only a NS-NS $B$ field.  These
backgrounds have received a great deal of attention recently, in the
cases where $K$ is $S^3\times T^4$ \cite{GKS,dBORT}, or $S^3\times
S^3\times S^1$ \cite{EFGT}, and have been studied in detail by using
perturbative methods.

Here we will initiate a study of the possible D-branes which can be
formulated consistently in a superstring background defined on
\begin{equation*}
\AdS_3 \times S^3 \times T^4~.
\end{equation*}
We will restrict ourselves to the case with a purely NS-NS B field.  This
will allow us to use the known superconformal theory (SCFT) underlying
this background, in order to apply the boundary state formalism
\cite{OOY, BBMOOY, KO, SDKS, RS, FuSch, STDNW, AS}.
 
The paper is organised as follows.  In Section~\ref{sec:background} we
will start with a short summary describing the bosonic background, in
order to set the notation and exhibit the conformal structure.  In
Section~\ref{sec:boundary} we discuss the boundary state formalism
adapted to this particular model.  We consider two classes of gluing
conditions: they both satisfy the basic requirement of conformal
invariance, but are distinguished by the amount of the bulk symmetry
they preserve.  In Section~\ref{sec:Dgeometry} we analyse the geometry
of the resulting D-brane configurations.  We consider the boundary
conditions that each of the two classes of gluing conditions give rise
to, and determine the types of D-branes they describe.  The
configurations described by gluing conditions which preserve the
infinite-dimensional symmetry underlying the WZW model generically
describe conjugacy classes translated by elements of the group.  This
result agrees in part with the similar analysis in \cite{AS}.  The
other type of gluing conditions gives rise to more complicated D-brane
configurations which include open submanifolds of dimension equal to
the dimension of the target manifold, subgroups and cosets; although a
direct comparison appears to be rather difficult, the two classes of
D-brane configurations obtained here bear a certain similarity to the
ones obtained in \cite{KlS} by analysing the open string WZW model.

In Section~\ref{sec:sigmamodel} we briefly discuss the classical
boundary conditions produced by the corresponding open string sigma
model in order to investigate the possibility of having D-branes whose
worldvolume fills the entire target space.  We find that this depends
crucially on whether the corresponding group is compact or not.

In Section~\ref{sec:susyext} we extend our analysis to the
supersymmetric case, and determine which D-brane configurations admit
an $N{=}1$ supersymmetric generalisation.  In
Section~\ref{sec:spacetimesusy} we discuss the fraction of spacetime
supersymmetry preserved by these configurations.   We end with a brief
discussion and summary of results, comparing our results with known
D-brane configurations in other $\AdS$ backgrounds.  The paper also
contains an appendix (written with JM~Figueroa-O'Farrill) in which we
investigate the conjugacy classes of $\SL(2,\R)$.

\section{The bosonic $\AdS_3\times S^3\times T^4$ background}
\label{sec:background}

The background we are interested in appears in type IIB string theory
on $\R^4\times\R^2\times T^4$ as the near-horizon limit of a system of
$p$ fundamental strings stretched along the $\R^2$ factor and $k$
NS-fivebranes stretched along the same $\R^2$ and wrapped on the
$T^4$.  Its metric corresponds to an exact string background
consisting of a flat space piece corresponding to $T^4$, a level $k$
$\SU(2)$ WZW theory and a level $k$ $\SL(2,\R)$ WZW theory.  We
therefore start with a WZW model having as the target space a direct
product of group manifolds $\mathbf{G}=\mathbf{G_1}
\times\mathbf{G_2}$, where $\mathbf{G_1}=\SL(2,\R)$,
$\mathbf{G_2}=\SU(2)$.  The corresponding WZW action will therefore be
a sum of two independent terms
\begin{equation}
I = I_{\SL(2,\R)}[g_1] + I_{\SU(2)}[g_2]~,\label{eq:wzw}
\end{equation}
where each term is of the form
\begin{equation*}
\frac{2}{k}I[g_i] = \int_{\Sigma} \langle g_i^{-1}\d g_i, g_i^{-1}\bar\d
                g_i\rangle + {\textstyle \frac{1}{6} }\int_B \langle
                g_i^{-1}dg_i, [g_i^{-1}dg_i,g_i^{-1}dg_i]\rangle~.
\end{equation*}
The full action of the theory includes, of course, a term
corresponding to $T^4$ and describing four commuting (compact) bosons.
Each of the fields $g_i$ is a map from a closed orientable Riemann
surface $\Sigma$ to the Lie group $\mathbf{G_i}$, $i=1,2$.  We denote
by $\gg=\gg_1\oplus\gg_2$ the corresponding Lie algebra, where
$\gg_1=\sl(2,\R)$ and $\gg_2=\su(2)$.  For these
algebras we choose the following bases of generators: $\{X_a\}$ for
$\gg_1$, and $\{Y_a\}$ for $\gg_2$ satisfying
\begin{equation*}\label{eq:gg1}
[X_1, X_2] = X_3~,\qquad [X_2, X_3] =- X_1~,\qquad [X_3, X_1] =- X_2~,
\end{equation*}
and
\begin{equation*}\label{eq:gg2}
[Y_1, Y_2] = Y_3~,\qquad [Y_2, Y_3] = Y_1~,\qquad [Y_3, Y_1] = Y_2~.
\end{equation*}
We also need to specify an invariant metric on $\gg$, which has a
diagonal form
\begin{equation*}\label{eq:eta}
\eta = \begin{pmatrix}
           \eta_1 & 0\\
           0 & \eta_2
         \end{pmatrix}~,
\end{equation*}
with components $(\eta_1)_{ab}\equiv \langle X_a,X_b\rangle =
\diag(+,+,-)$ and $(\eta_2)_{ab}\equiv \langle Y_a,Y_b\rangle =
\diag(+,+,+)$.

We will use the following parametrisation\footnote{This
  parametrisation of $\AdS_3$ is different from the one given by the
  Gauss decomposition,
\begin{equation*}
g_1 = e^{\bar\gamma (X_2-X_3)} e^{2r X_1}e^{\gamma (X_2+X_3)}~, 
\end{equation*}
and which yields the familiar metric $ds_1^2=\frac{1}{u^2}du^2 + u^2
d\gamma d\bar\gamma$, where $u=e^r$.}  for the group manifold
$\mathbf{G}$:
\begin{equation}\label{eq:parg}
g_1 = e^{\theta_2 X_2}e^{\theta_1 X_1}e^{\theta_3 X_3}~,\qquad 
g_2 = e^{\phi_2 Y_2}e^{\phi_1 Y_1}e^{\phi_3 Y_3}~,
\end{equation}
where $\theta_{\mu}$ and $\phi_{\mu}$, $\mu=1,2,3$, play the r\^ole of
the spacetime fields.  In terms of them \eqref{eq:wzw} becomes a
sigma-model action, with the spacetime metric and 2-form given by
\begin{equation*}\label{eq:met1}
ds_1^2 = d\theta_1 d\theta_1 + d\theta_2 d\theta_2 - d\theta_3
d\theta_3 + 2\sinh\theta_1 d\theta_2 d\theta_3~,
\end{equation*}
\begin{equation*} 
d B_1 = - \cosh\theta_1 d\theta_1 d\theta_2 d\theta_3~,\label{eq:B1}
\end{equation*}
on $\SL(2,\R)$, and 
\begin{equation*}\label{eq:met2}
ds_2^2 = d\phi_1 d\phi_1 + d\phi_2 d\phi_2 + d\phi_3 d\phi_3 -
2\sin\phi_1 d\phi_2 d\phi_3~, 
\end{equation*}
\begin{equation*} 
d B_2 = \cos\phi_1 d\phi_1 d\phi_2 d\phi_3~,\label{eq:B2}
\end{equation*}
for $\SU(2)$.  (The torus will have of course flat metric and no $B$
field.)

The exact conformal invariance of this model is based, as is well
known, on its infinite-dimensional symmetry group
$\mathbf{G}(z)\times\mathbf{G}(\bar z)$ characterised by the conserved
currents $I(z)=-k\d g_1 g_1^{-1}$ and $\bar I(\bar z)=kg_1^{-1}\bar\d
g_1$ corresponding to $\sl(2,\R)$, and $J(z)=-k\d g_2
g_2^{-1}$ and $\bar J(\bar z)=kg_2^{-1}\bar\d g_2$ corresponding to
$\su(2)$.  These currents generate an affine Lie algebra
$\widehat\gg_1\oplus\widehat\gg_2$, with $\widehat\gg_1$ described by
\begin{equation}
I_a (z)I_b (w) = \frac{k(\eta_1)_{ab}}{(z-w)^2} + \frac{{f_{ab}}^c
                 I_c (w)}{z-w} + \reg~,\label{eq:affg1}
\end{equation}
where the parameter $k$ is related to the level $x$ of the
affine algebra by $k=x+g^*$, where $g^*$ is the dual Coxeter number.
Similarly, for $\widehat\gg_2$ we have
\begin{equation}
J_a (z)J_b (w) = \frac{k(\eta_2)_{ab}}{(z-w)^2} +
          \frac{{f_{ab}}^c J_c (w)}{z-w} + \reg~,\label{eq:affg2}
\end{equation}
whereas the free bosons on the torus, described by the fields
$\varphi_i$, $i=1,2,3,4$, satisfy the standard OPEs
\begin{equation}
  \d\varphi_i(z)\d\varphi_j(w) = \frac{\delta_{ij}}{(z-w)^2} +
\reg~,\label{eq:T4}
\end{equation}
with similar OPEs for the antiholomorphic sector.  The corresponding
CFT is then described by the energy-momentum tensor
\begin{equation*}
\sT = \Omega_1^{ab}(I_aI_b) + \Omega_2^{ab}(J_aJ_b) + 
      \sum_{i=1}^{4} (\d\varphi_i\d\varphi_i)~, 
\end{equation*}
where $\Omega_1^{ab}$ and $\Omega_2^{ab}$ are components of the
inverse of the following invariant metric
\begin{equation*}\label{eq:Omega}
\Omega = \begin{pmatrix}
           \Omega_1 &   0\\
                0   & \Omega_2
         \end{pmatrix}~,
\end{equation*}
with the components given by $\Omega_1=2(k+1)\eta_1$ and $\Omega_2 =
2(k-1)\eta_2$.  The central charge of this CFT is given by
\begin{equation*}
c = \frac{3k}{k+1} + \frac{3k}{k-1} + 4~.
\end{equation*}
The choice of equal levels for $\widehat\gg_1$ and $\widehat\gg_2$ was 
motivated by the fact that, in this case, the corresponding $N{=}1$
supersymmetric background is a critical superstring theory, as we will 
see in Section~\ref{sec:susyext}.

\section{Boundary states}
\label{sec:boundary}

The boundary state formalism (see, e.g., \cite{CLNY,PC,Li,CK}) has
become in the last years one of the main approaches to the study of
D-branes in type II string backgrounds \cite{OOY, BBMOOY, KO, SDKS,
  RS, FuSch}.  Using an explicit knowledge of the CFT underlying a
given string background, one describes a D-brane configuration through
a set of boundary conditions, relating the left-- and the right--
moving conformal structures in such a way that conformal invariance is
preserved.  In the case of a background described by a WZW model the
fields in terms of which the conformal structure of the model is
realised are the affine currents.  It is therefore convenient to
impose the boundary conditions on these currents in order to have
under control the conformal invariance of the resulting
configurations.

In what follows we will consider two different classes of
gluing\footnote{In order to avoid confusion between the boundary
  conditions satisfied by the chiral currents, $J$ and $\bar J$, and
  the ones satisfied the field $g$ or its components, we will refer to
  the boundary conditions on the chiral currents as `gluing
  conditions', reserving the term of boundary conditions to the fields
  themselves.}  conditions which give rise to different configurations
of D-branes.  Both of them will be defined in terms of a Lie algebra
automorphism, $R:\gg\to\gg$, which preserves the metric $\eta$; in
other words we have
\begin{equation}
[R(Z_a),R(Z_b)] = R([Z_a,Z_b])~,\label{eq:Rhom}
\end{equation}
\begin{equation}
R^T \eta R = \eta~,\label{eq:Rg}
\end{equation}
where $\{Z_a\}$ is a given basis in $\gg$, in terms of which $R$ is
given by $R(Z_a)=Z_b{R^b}_a$.  We can now define the two classes of
gluing conditions\footnote{We will be working here in the open string
  picture.} as being the following:
\begin{itemize}
\item[(i)] \emph{type-N gluing conditions}, which can be thought of as 
  a generalisation of the Neumann conditions, are given by
\begin{equation}
J_a(z) + {R^b}_a {\bar J}_b(\bar z) = 0~,\label{eq:Nbc} 
\end{equation}
at the boundary of the worldsheet.  Notice that these gluing
conditions do not preserve the infinite--dimensional affine symmetry
of the current algebra (correcting a statemnet made in \cite{STDNW});

\item[(ii)] \emph{type-D gluing conditions}, which can be thought of
  as a generalisation of Dirichlet conditions, read
\begin{equation}
J_a(z) - {R^b}_a {\bar J}_b(\bar z) = 0~.\label{eq:Dbc} 
\end{equation}
By contrast, this type of gluing conditions does preserve the current
algebra of the bulk theory.
\end{itemize}

These gluing conditions have to satisfy the basic consistency
requirement, which is conformal invariance.  This means that the
holomorphic and the antiholomorphic sectors are related by an
automorphism of the corresponding CFT.  In the bosonic case, since the
automorphism group of the Virasoro algebra is trivial, we impose
\begin{equation*}
\sT(z) = \bar\sT(\bar z)~,\label{eq:ci}
\end{equation*}
at the boundary.  In this case, the requirement of conformal
invariance translates into the condition
\begin{equation}
R^T \Omega R = \Omega~,\label{eq:Rmet}
\end{equation}
for either of the two types of gluing conditions.

Ignoring the flat part of the target space---that is, $T^4$---for
which the possible D-branes are known, we take $\gg$ to be the direct
sum of the two Lie algebras $\gg_1 = \sl(2,\R)$ and
$\gg_2=\su(2)$.  Because these Lie algebras are different
real forms of the same complex Lie algebra $\sl(2,\C)$,
there is no nontrivial homomorphism between them.  This implies that
the matrix of gluing conditions $R$, defined by the automorphism
$R:\gg_1\oplus\gg_2\to\gg_1\oplus\gg_2$, must take a block-diagonal
form
\begin{equation}\label{eq:R}
R = \begin{pmatrix}
           R_1 &   0\\
           0   & R_2
         \end{pmatrix}~,
\end{equation}
where $R_1:\gg_1\to\gg_1$ and $R_2:\gg_2\to\gg_2$.  Then from
\eqref{eq:Rmet} and \eqref{eq:Rg} we deduce that $R_1$ and $R_2$ must
separately preserve the metric on $\SL(2,\R)$ and $\SU(2)$
respectively, and therefore $R_1$ defines an element of
$\mathrm{O}(2,1)$, whereas $R_2$ is an element of $\mathrm{O}(3)$.

On the other hand, from \eqref{eq:Rhom} it follows that both $R_1$
and $R_2$ are Lie algebra automorphisms, corresponding to
$\sl(2,\R)$ and $\su(2)$, respectively.
Explicitly, each of these two automorphism conditions translates into
a condition on the corresponding matrix, that is
\begin{equation*}
\det(R_1) = \det(R_2) = 1~,
\end{equation*}
which makes $R_1$ belong to $\SO(2,1)$, and $R_2$ to $\SO(3)$.  

These results can be summarised as follows.  We have identified two
classes of gluing conditions on the group manifold $\SL(2,\R)\times
\SU(2)$ which preserve conformal invariance.  Each of them is
described in terms of an automorphism of the Lie algebra
$\sl(2,\R)\oplus\su(2)$, but only the type-D
gluing conditions \eqref{eq:Dbc} preserve also the
infinite-dimensional symmetry of the current algebra of the bulk
theory.  In what follows we will analyse in detail both these types of
configurations in order to identify the D-branes they describe.

\section{D-brane solutions and geometry}
\label{sec:Dgeometry}

The geometric interpretation of the gluing conditions defined on the
chiral currents of the WZW theory in terms of D-brane configurations
is arguably the most subtle issue of this approach.  The precise
statement of the problem is the following: given a set of gluing
conditions for the chiral currents and a fixed but otherwise arbitrary
point $g$ in the target group manifold, find the possible D-branes
which pass through $g$ and are described by these gluing conditions.

The difficulty lies with the fact that the gluing conditions imposed
on the affine currents, despite being a natural nonabelian
generalisation of the boundary conditions in a free theory are
\emph{not}, strictly speaking, boundary conditions.  The flat space
boundary conditions are defined in the tangent space of the target
manifold and therefore the eigenvalues and eigenvectors of the
corresponding matrix $R$ identify the Neumann and Dirichlet
directions.  In the group manifold case the gluing conditions take
values in the tangent space of $\mathbf{G}$ \emph{at the identity},
that is $T_e\mathbf{G}\equiv\gg$, because the currents themselves are
Lie algebra valued objects.  Hence, in order to interpret
geometrically the algebraic gluing conditions we must first of all
`translate' them into boundary conditions in $T_g\mathbf{G}$, and then
determine what the Neumann and Dirichlet directions are in this case.

In case of the $\AdS_3\times S^3$ background, the matrix $R$ has a
block diagonal form \eqref{eq:R}, which allows us to analyse the
D-brane configurations on $\AdS_3$ and $S^3$ separately.  Thus $R_1$,
which is an element of $\SO(2,1)$ can be thought of as a Lorentz
transformation in $\mathbb{R}^{2,1}$.  In three dimensions, any
Lorentz transformation leaves a vector invariant.  Depending on the
causal type of this vector, we can distinguish three types of Lorentz
transformations:
\begin{itemize}
\item[(i)] \emph{spatial rotations}, typically of the form
\begin{equation}
R_1 = \begin{pmatrix}\label{eq:R1i}
       \cos\alpha & \sin\alpha & 0\\
      -\sin\alpha & \cos\alpha & 0\\
       0 & 0 & 1
      \end{pmatrix}~,\qquad\alpha\in\mathbb{R}~.
\end{equation}
In this case $R_1$ has generically one $+1$ eigenvalue, with the
corresponding eigenvector being time-like;
\item[(ii)] \emph{boosts}, typically of the form
\begin{equation}
R_1 = \begin{pmatrix}\label{eq:R1ii}
       \cosh\alpha & 0 & \sinh\alpha\\
       0 & 1 & 0\\
       \sinh\alpha & 0 & \cosh\alpha\\
      \end{pmatrix}~,\qquad\alpha\in\mathbb{R}~.
\end{equation}
In this case $R_1$ has generically one $+1$ eigenvalue, but with the
corresponding eigenvector being space-like;
\item[(iii)] \emph{null rotations}, typically of the form
\begin{equation}
R_1 = \begin{pmatrix}\label{eq:R1iii}
       1 & -a & a\\
       a & 1-\half a^2 & \half a^2\\
       a & -\half a^2 & 1+\half a^2\\
      \end{pmatrix}~,\qquad a\in\mathbb{R}~.
\end{equation}
In this case $R_1$ has generically a $+1$ eigenvalue whose eigenvector 
is light-like.  In contrast with the other two cases, a null rotation
is not diagonalisable over the complex numbers.  This makes the
geometric interpretation of the corresponding configuration relatively
difficult.
\end{itemize}

By contrast, a D-brane configuration in $\SU(2)$ is described by an
element $R_2$ in $\SO(3)$ that is, a rotation in $\R^3$, typically of
the form
\begin{equation}
R_2 = \begin{pmatrix}\label{eq:R2}
       \cos\beta & \sin\beta & 0\\
      -\sin\beta & \cos\beta & 0\\
       0 & 0 & 1                
      \end{pmatrix}~,\qquad\beta\in\mathbb{R}~.
\end{equation}
It is generically characterised by a $+1$ eigenvalue, whose
eigenvector corresponds to the direction which is left invariant by
the rotation.

\subsection*{Type-N configurations}

We now arrive at the basic fact on which our geometric interpretation
of the gluing conditions is based.  Here we will only state this
result (details will appear elsewhere \cite{SDnotes}) and proceed to
apply it to the case of the group manifold corresponding to
$\AdS_3\times S^3$.  

Let us parametrise the group manifold $\mathbf{G}$ by introducing the
coordinates $X^{\mu}$, with $\mu=1,...,\dim\mathbf{G}$; we also
introduce the left- and right-invariant vielbeins, $e$ and $\bar e$,
defined by
\begin{equation*}
g^{-1}dg = {e^a}_{\mu}~dX^{\mu}Z_a~,\qquad\qquad
dg g^{-1} = {\bar e}^a{}_{\mu}~dX^{\mu}Z_a~.
\end{equation*}
The gluing conditions \eqref{eq:Nbc} will then give rise to
the following boundary conditions at a generic point $g$ in
$\mathbf{G}$:
\begin{equation*}
\d X^{\mu} = {\tilde R(g)}^{\mu}{}_{\nu} \bar\d X^{\nu}~,
\end{equation*}
where the matrix of boundary conditions $\tilde R(g)$ is given by
\begin{equation*}
\tilde R(g) = {\bar e}^{-1} R e~.
\end{equation*}
Notice that $\tilde R(g)$, which describes the boundary conditions at
a given point in the target space, depends on that point through the
invariant vielbeins.  One can now identify the Neumann and Dirichlet
directions by analogy with the flat space case.  At a given point, a
Dirichlet boundary condition corresponds to a $-1$ eigenvalue of the
matrix $\tilde R(g)$, which means that the directions normal to the
worldvolume of the D-brane are spanned by the corresponding
eigenvectors of $\tilde R(g)$.  All the other eigenvalues describe
Neumann boundary conditions (in the presence of a $B$ field) and the
corresponding eigenvectors span the tangent space of the worldvolume
of the D-brane.

We now start with $\SL(2,\R)$, and consider the simplest possible case
where $R_1=\1$.  In this case the boundary conditions at a point
$g_1(\theta_{\mu})$ will read
\begin{equation*}
\d\theta^{\mu} = {\tilde R_1}(g_1)^{\mu}{}_{\nu}\bar\d\theta^{\nu}~,
\end{equation*}
where $\tilde R_1(g_1) = \bar e_1^{-1}e_1$.  If we evaluate the matrix
of boundary conditions at the identity we obtain $\tilde R_1(e)=\1$,
which indicates that we have three 
Neumann directions spanning the whole $\sl(2,\R)$.  In other
words, the identity in $\SL(2,\R)$ belongs to a D2-brane.  

If we now move away from the identity, $\tilde R_1$ will no longer be
$\1$.  Instead, we obtain that $\tilde R_1$ always has one $+1$
eigenvalue and two complex conjugate eigenvalues.  Hence, at generic
points in the group manifold, $\tilde R_1(g_1)$ will still give rise
to three Neumann directions; however there will be a submanifold of
$\SL(2,\R)$ where $\tilde R_1(g_1)$ will have at least one $-1$
eigenvalue (two, in fact, since $\det\tilde R_1=1$).  In other words,
at each point on this particular submanifold we will have one Neumann
and two Dirichlet directions.  This submanifold can be described as
the zero locus $\eF_1$ of a function, $F_1(g_1)\equiv\Tr\tilde
R_1(g_1) + 1$, which in our parametrisation is given by
\begin{multline*}
  F_1 = 1 + \cosh\theta_1\cosh\theta_2 + \cosh\theta_2\cos\theta_3 +
  \cos\theta_3\cosh\theta_1\\
  + \sinh\theta_1\sinh\theta_2\sin\theta_3~.
\end{multline*}
One can show that $F_1$ is a class function; that is, $F_1(h^{-1}gh) = 
F_1(g)$ for all group elements $g,h$.  Indeed, if we consider the
vector fields that generate the adjoint action of the group
$\SL(2,\R)$
\begin{equation}
  \label{H_a}
  H_a(g_1) = \left( ({\bar e_1}^{-1})^{\mu}{}_a - (e_1^{-1})^{\mu}{}_a
  \right) \d_{\mu}~,\qquad\qquad a=1,2,3,
\end{equation}
we can check that $F_1$ is annihilated by them, that is
\begin{equation*}
  H_a(g_1)\cdot F_1(g_1) = 0~,
\end{equation*}
for all $a=1,2,3$.  Hence we have that $\eF_1$ consists of adjoint
orbits---that is, conjugacy classes.

At every point in $\eF_1$ we have one Neumann and two Dirichlet
boundary conditions.  The vector field corresponding to the Neumann
direction therefore spans the worldline of a D-particle.  In order for
this picture to be consistent we must verify that the worldlines of
these D0-branes lie within $\eF_1$.  Indeed, a straightforward
calculation shows that we have
\begin{equation*}
\left. V_1(g_1)\cdot F_1(g_1)\right|_{F_1(g_1)=0} = 0~,
\end{equation*}
which allows us to conclude that $V_1$ is tangent to $\eF_1$.

What happens now if we start with a matrix $R_1$ which is different
from the identity?  First of all we recall that $\sl(2,\R)$ is a
simple Lie algebra, and therefore $R_1$ is an inner automorphism;
hence it can be identified with $\Ad_{r_1}$, for some group element
$r_1$.  As we will see in the next paragraph, the effect of an inner
automorphism at the level of the gluing conditions is a translation in
the group manifold.  More precisely, the submanifold on which $\tilde
R_1=\bar e_1^{-1}R_1 e_1^{-1}$ has $-1$ eigenvalues, which we denote
by $\eF_1^{(r_1)}$, is the zero locus of a function $F_1^{(r_1)}$
which satisfies $F^{(r_1)}(g)=F(gr_1^{-1})$.  Thus $\eF_1^{(r_1)}$ is
nothing but the translation of the previous $\eF_1$ by the group
element $r_1$, that is $\eF_1^{(r_1)}=\eF_1 r_1$.  Hence through every
point in $\eF_1 r_1$ passes a D0-brane whose worldline lies on $\eF_1
r_1$.  Moreover, the Neumann eigenvectors tangent to the worldline of
the D0-branes in $\eF_1 r_1$ can be obtained by translating
accordingly the corresponding Neumann eigenvectors tangent to the
D0-branes in $\eF_1$.


A particularly interesting case is the one where $R_1$ itself has two
$-1$ eigenvalues (this can be obtained by taking $R_1$ of the form
\eqref{eq:R1i} with $\alpha=\pi$).  In this case, the corresponding
surface $\eF_1 r_1$ passes through the identity element in
$\SL(2,\R)$, and therefore there exists a particular D0-brane whose
worldline passes through the identity.  Its tangent vector takes a
particularly simple form at the identity, being given by
$V_1=\d_{\theta_3}$, as expected.  The worldline of this D0-brane is
nothing but the subgroup of $\SL(2,\R)$ generated by $X_3$.  Moreover,
the translation of this particular solution gives rise to D0-brane
configurations whose worldlines are cosets in $\SL(2,\R)$.

We can now analyse the $\SU(2)$ case in a similar fashion.  If we
start with $R_2=\1$, the boundary conditions at a point $g_2(\phi_i)$
will read
\begin{equation*}
\d\phi^{\mu} = {\tilde R_2}(g_2)^{\mu}{}_{\nu}\bar\d\phi^{\nu}~,
\end{equation*}
where $\tilde R_2(g_2) = \bar e_2^{-1}e_2$.  At the identity we have
$\tilde R_2(e)=\1$, which gives three Neumann directions spanning
$\su(2)$.  We can therefore conclude that the identity in
$\SU(2)$ belongs to an euclidean D2-brane.  Away from the identity,
$\tilde R_2$ will have one $+1$ and two complex conjugate eigenvalues,
which generically describe three Neumann directions.  Similarly
to the previous case, there will be a submanifold $\eF_2$ of $\SU(2)$
where $\tilde R_2$ has two $-1$ eigenvalues.  This submanifold can be
described as the zero locus of the function $F_2(g_2)\equiv\Tr\tilde
R_2(g_2) + 1$ which reads
\begin{multline*}
  F_2 = 1 + \cos\phi_1\cos\phi_2 + \cos\phi_2\cos\phi_3 +
  \cos\phi_3\cos\phi_1\\
  + \sin\phi_1\sin\phi_2\sin\phi_3~.
\end{multline*}
As before, one can show that $F_2$ is a class function,
and hence that $\eF_2$ consists of conjugacy classes.

One can show, along the same lines, that $\eF_2$ is foliated by the
worldlines of the euclidean D0-branes whose tangent vectors are given
by the eigenvectors $V_2$ of $\tilde R_2$ corresponding to the $+1$
eigenvalue.

Finally, if instead of $R_2=\1$ we take $R_2=\Ad_{r_2}$ (since, as
before, $R_2$ is an inner automorphism) the resulting D-brane
configurations can be understood as translations in the group manifold
of the ones obtained in the case $R_2=\1$.  Thus, we obtain in
particular euclidean D0-branes whose worldlines lie in $\eF_2 r_2$.

\subsection*{Type-D configurations}

This type of gluing gluing conditions \eqref{eq:Dbc} have been
recently analysed in \cite{AS}, where the resulting D-brane
configurations have been identified with conjugacy classes.  Here
however, by using a slightly different point of view, we will arrive
at partially different conclusions.

As we mentioned before, the reason for which the gluing conditions
\eqref{eq:Dbc} cannot immediately be interpreted as boundary
conditions in the target space is the fact that the currents
themselves are Lie algebra valued objects. 
In order to obtain a boundary condition from the gluing condition
\eqref{eq:Dbc}, we must translate the currents, which take values in
$\gg$, into objects taking values in $T_g\mathbf{G}$.  In this way one
obtains \cite{SDnotes}
\begin{equation}
\d g = \mathbf{R}(g) \bar\d g~,\label{eq:dbc}
\end{equation}
with the matrix of boundary conditions given by 
\begin{equation*}
\mathbf{R}(g) = - (\rho_g)_* \circ R \circ (\lambda_g)_*^{-1}~,
\end{equation*}
where $\lambda_g$ and $\rho_g$ stand for left- and
right-multiplication by $g$ in $\mathbf{G}$.  The condition
\eqref{eq:dbc} now takes place in $T_g\mathbf{G}$, and it is the
corresponding, point-dependent matrix $\mathbf{R}(g)$ which determines
the Neumann and Dirichlet directions.  Indeed, at a given point $g$ in
$\mathbf{G}$, a Dirichlet boundary condition corresponds to a $-1$
eigenvalue of $\mathbf{R}(g)$, which means that the directions normal
to the worldvolume of the D-brane are spanned by the corresponding
eigenvectors of $\mathbf{R}(g)$.  All the other eigenvalues describe
Neumann boundary conditions and the corresponding eigenvectors span
the tangent space of the worldvolume of the D-brane.
 
If $R$ is taken to be the identity matrix, then $\mathbf{R}(g) = -
\Ad_{g^{-1}}$, and the corresponding D-branes can be identified with
the conjugacy classes of the group $\mathbf{G}$ \cite{AS}.  Indeed, in
this case, and provided that the metric $G$ restricts nondegenerately
to the conjugacy class $C$ of $g$, the tangent space at $g$ splits
into the tangent space to the conjugacy class and its perpendicular
complement, which can be identified with the tangent space to the
centraliser subgroup $Z$ of $g$:
\begin{equation*}
T_g\mathbf{G} = T_g C \oplus T_g Z\quad\text{with}\quad T_gC \perp T_g 
Z~.
\end{equation*}
Moreover $\Ad_{g^{-1}}$ restricts to the identity on $T_gZ$, which
means that the Dirichlet directions span $T_g Z$.  Furthermore, the
Neumann directions span $T_g C$, and hence the worldvolume of the
D-brane can be identified with $C$.

Let us now consider the case of an arbitrary $R$.  In our case, since
both $\sl(2,\R)$ and $\su(2)$ are simple Lie
algebras, we can restrict ourselves to the case where $R$ is an inner
automorphism, and hence can be identified with $\Ad_r$, for some group
element $r$.  Therefore the corresponding boundary conditions can be
written in the following form
\begin{equation*}
\d\tilde g = -\Ad_{\tilde g^{-1}} \bar\d\tilde g~,
\end{equation*}
with $\tilde g = gr^{-1}$.  This implies that the corresponding
D-brane lies along the right--translate $Cr$ of the conjugacy class of
$g$ by the element $r$.  This result contradicts the statement made in
\cite{AS} according to which inner automorphisms, being a ``symmetry
of the model'' cannot result in D-brane configurations different from
the the ones already described by $R=\1$.  Although inner
automorphisms are symmetries of the background, they are not
necessarily symmetries the theory containing a D-brane.  This fact is
not surprising, as D-branes break some of the bulk symmetries even in
flat space (e.g., translational symmetry).

We can now specialise to our particular background where, as usual, we
will consider the two groups separately.  The conjugacy classes of
$\SU(2)$ are well known, and have been recently discussed in
\cite{AS}.  We have listed them in Table~\ref{tab:SU(2)conj-classes}.
They are parametrised by $S^1/\Z_2$, which we can understand as the
interval $\theta\in[0,\pi]$.  The conjugacy classes corresponding to
$\theta=0,\pi$ are points, corresponding to the elements $\pm e$ in
the centre of $\SU(2)$, whereas the classes corresponding to
$\theta\in(0,\pi)$ are spheres.  If we picture $\SU(2)$, which is
homeomorphic to the 3-sphere, as the one-point compactification of
$\R^3$ where the sphere at infinity is collapsed to a point, the
foliation of $\SU(2)$ by its conjugacy classes coincides with the
standard foliation of $\R^3$ by 2-spheres with two degenerate spheres
at the origin and at infinity.

\begin{table}[h!]
\centering
\setlength{\extrarowheight}{3pt}
\begin{tabular}{|>{$}c<{$}|>{$}c<{$}|>{$}c<{$}|}\hline
\text{Class} & \text{Element} & \text{Topology} \\
\hline\hline
\eC'_e & \begin{pmatrix} 1 & 0\\ 0 & 1 \end{pmatrix} & \text{point} \\
\eC'_{-e} & \begin{pmatrix} -1 & 0\\ 0 & -1 \end{pmatrix} & \text{point} \\
\eC'_\theta & \begin{pmatrix} e^{i\theta} & 0\\ 0 & e^{-i\theta}
             \end{pmatrix} & S^2 \\ 
\hline
\end{tabular}
\vspace{8pt}
\caption{$\SU(2)$ conjugacy classes}
\label{tab:SU(2)conj-classes}
\end{table}

Let us now turn to $\SL(2,\R)$.  Its conjugacy classes are computed in 
the appendix and can be labelled by eight types of $2\times 2$
matrices.  Those classes which are metrically nondegenerate can be
interpreted as D-branes and are listed in Table~\ref{tab:sl2rdbranes}.
As in the case of $\SU(2)$ we have two point-like D-branes
corresponding to the two elements in the centre of $\SL(2,\R)$ as well
as a family of euclidean D-strings with planar topology, but in
addition there is also a family of D-strings with cylindrical
topology.

\begin{table}[h!]
\centering
\setlength{\extrarowheight}{3pt}
\begin{tabular}{|>{$}c<{$}|>{$}c<{$}|>{$}c<{$}|}\hline
\text{Class} & \text{Element} & \text{Topology} \\
\hline\hline
\eC_e & \begin{pmatrix} 1 & 0\\ 0 & 1 \end{pmatrix} & \text{point}\\
\eC_{-e} & \begin{pmatrix} -1 & 0\\ 0 & -1 \end{pmatrix} & \text{point}\\
\eC_\theta & \begin{pmatrix} \cos\theta & \sin\theta\\ -\sin\theta &
\cos\theta \end{pmatrix} & \R^2\\
\eC_\lambda & \begin{pmatrix} \lambda & 0\\ 0 & 1/\lambda \end{pmatrix} &
\R\times S^1\\
\hline
\end{tabular}
\vspace{8pt}
\caption{$\SL(2,\R)$ D-branes based on conjugacy classes}
\label{tab:sl2rdbranes}
\end{table}

Let us now summarise our findings so far.  We have seen that type--N
gluing conditions give rise to D5-, D3-, and D1-branes in
$\AdS_3\times S^3$ whose worldvolumes are of the form $N_1\times N_2$,
with $N_1$ and $N_2$ three-- or one--dimensional submanifolds of
$\AdS_3$ and $S^3$, respectively.  Moreover, particular solutions for
$N_1$ and $N_2$ include subgroups and cosets of $\SL(2,\R)$ and
$\SU(2)$, respectively.  By contrast, type--D gluing conditions, for a
given $R=\Ad_r$, describe D-branes whose ``worldvolumes'' are shifted
conjugacy classes of the form $\eC r_1\times\eC'r_2$, which can be
$0$-, $2$- or $4$-dimensional.

\section{Relation to the sigma model approach}
\label{sec:sigmamodel}

One of the most surprising results of the boundary state analysis of
the possible D-branes in $\AdS_3\times S^3$ is the absence of D-brane
configurations which fill the entire group manifold.  In this section
we pause for a moment our analysis via the boundary state approach to
consider the classical sigma model which corresponds to our WZW
theory.  Our main aim here is to investigate the possibility of having
D-branes which fill the whole target space, that is $\AdS_3\times
S^3$.

The action of a generic WZW model on a 2-space with a disc topology
with an additional interaction (1-form field $A$) at the boundary
reads
\begin{equation}
S = \int_{\Sigma} \langle g^{-1}\d g, g^{-1}\bar\d g\rangle + 
    \int_{\Sigma} g^* B + \int_{\d\Sigma} g^* A~.\label{eq:owzw}  
\end{equation}
Here the worldsheet $\Sigma$ is a two-dimensional manifold with
boundary $\d\Sigma$, and $B$ represents a particular choice for the
antisymmetric tensor field.  A D-brane configuration is characterised
in this setting (for more details see \cite{KlS}) by a two-form
$\alpha$ living on the worldvolume $D$ of the D-brane (in which the
boundary of the string worldsheet $\d\Sigma$ is included), and
satisfying $d\alpha=\left.dB\right|_D$.  Since
$\left.d(B-\alpha)\right|_D=0$, one can define locally the one-form
potential $A$ such that $dA=B-\alpha$.  $S$ may be viewed as a special
case of an action for an open string propagating on a group manifold
and coupled to $A$ at the boundary.  In the case of $\AdS_3\times S^3$
the action consists of two independent terms, $S_1+S_2$, such that
each of them is of the form \eqref{eq:owzw}, only with different
target spaces, corresponding to the two groups, $\SL(2,\R)$ and
$\SU(2)$, respectively.  However for the most part we will work with
the generic form of the action \eqref{eq:owzw}, and we will consider
the two components separately only at the very end.

The infinitesimal variation of $S$ contains a bulk term (yielding the
same equations of motion as in the closed string case) and a boundary
term which reads
\begin{equation*}
\left. \int_{\d\Sigma} d\tau (g^{-1}\delta g)^a \left[ \eta_{ab} (g^{-1}
\d_{\sigma} g)^b -i \alpha_{ab} (g^{-1}\d_{\tau}
g)^b\right]\right|^{\sigma=\pi}_{\sigma=0}
\end{equation*}
\begin{equation*}
=\left.\int_{\d\Sigma}d\tau \delta X^{\mu}
p_{\mu}\right|^{\sigma=\pi}_{\sigma=0}~,
\end{equation*}
where $\eta$ is the generic metric on the group manifold and $X^{\mu}$
are the coordinates introduced in the previous section.  We have
denoted by $p_{\mu}$ the component of the 2-momentum normal to the
boundary $\d\Sigma$ which is given by
\begin{equation*}
p_{\mu} = G_{\mu\nu}\d_{\sigma}x^{\nu} -i\alpha_{\mu\nu}
          \d_{\tau}x^{\nu}~, 
\end{equation*}
where $G_{\mu\nu} = {e^a}_{\mu} \eta_{ab}{e^b}_{\nu}$, $\alpha_{\mu\nu} = 
{e^a}_{\mu} \alpha_{ab}{e^b}_{\nu}$.

Having Neumann boundary conditions in all directions amounts to
imposing $p_{\mu}|_{\d\Sigma}=0$, for all $\mu$.  In order to compare
these conditions with the boundary conditions \eqref{eq:Nbc}, we must
express them in terms of the same quantities---that is, in terms of
the conserved currents.  For this we use the expressions of the
currents in terms of the spacetime fields
\begin{equation} \label{eq:J(x)}
J_a = -\eta_{ab}{\bar e}^b{}_{\mu} \d X^{\mu}~, \qquad 
\bar J_a = \eta_{ab}{e^b}_{\mu} \bar\d X^{\mu}~,
\end{equation}
in order to rewrite $p_{\mu}$ as follows:
\begin{equation*}
p_{\mu} = -[\delta_{\mu}{}^{\rho} - \alpha_{\mu\nu}G^{\nu\rho}]{\bar
           e}^a{}_{\rho} J_a - [\delta_{\mu}{}^{\rho} +
          \alpha_{\mu\nu}G^{\nu\rho}]{e^a}_{\rho} \bar J_a~.
\end{equation*}
Then the Neumann boundary conditions take the following form
\begin{equation}
  \label{eq:cBC}
  J + M \bar J = 0~,
\end{equation}
where the matrix $M$ depends on the background fields, being given by 
\begin{equation}
  \label{eq:cBCN}
  M \equiv \bar e^{-T}{\frac{\1 + \alpha G^{-1}}{\1 -\alpha G^{-1}}}e^T~.
\end{equation}
Since we imposed Neumann conditions in all directions, it is natural
to expect that we obtain a D-brane whose worldvolume has the same
dimension as the dimension of the target space.  Notice however that
from this we cannot immediately deduce that the D-brane worldvolume
literally fills the whole target space.  We will return to this point
at the end of this section.

In our case, since all the relevant quantities (that is, the
background fields and the corresponding vielbeins) take a
block-diagonal form with respect to the two group components, the
matrix $M$ will do so as well,
\begin{equation*}\label{eq:M}
M = \begin{pmatrix}
           M_1 &   0\\
           0   & M_2
         \end{pmatrix}~.
\end{equation*}
Therefore we can compute the two components separately.

\subsection{The $\SL(2,\R)$ component}

We can now compute the matrix of boundary conditions $M_1$
corresponding to the open WZW action $S_1$ with target space
$\SL(2,\R)$.  In order to do this we use the parametrisation of
$\SL(2,\R)$ given by the first expression in \eqref{eq:parg}.  Then
the invariant vielbeins $e_1$ and $\bar e_2$ are given by
\begin{gather*}
e_1 = \begin{pmatrix}
      \cos\theta_3 & -\cosh\theta_1\sin\theta_3 & 0\\
      \sin\theta_3 & \cosh\theta_1\cos\theta_3 & 0\\
       0   &    -\sinh\theta_1     & 1
    \end{pmatrix}\\
\bar e_1 = \begin{pmatrix}
         \cosh\theta_2 & 0 & -\cosh\theta_1\sinh\theta_2\\
                     0 & 1 & \sinh\theta_1\\
        -\sinh\theta_2 & 0 & \cosh\theta_1\cosh\theta_2
         \end{pmatrix}.
\end{gather*}
The corresponding background metric will be $(G_1)_{\mu\nu} =
{e^a}_{\mu} (\eta_1)_{ab}{e^b}_{\nu}$.  If we choose $\alpha_1 =
\sinh\theta_1 d\theta_2 d\theta_3$ we find that the matrix of the
boundary conditions is given by
\begin{multline*}
\small
M_1 = \begin{pmatrix}
                \cosh\theta_2 & 0 & \sinh\theta_2\\
                            0 & 1 & 0\\
                \sinh\theta_2 & 0 & \cosh\theta_2
              \end{pmatrix}
\begin{pmatrix}
                1 &             0 & 0\\
                0 & \cosh\theta_1 & \sinh\theta_1\\
                0 & \sinh\theta_1 & \cosh\theta_1 
              \end{pmatrix}\\
\times\small
\begin{pmatrix}
                \cos\theta_3 & \sin\theta_3 & 0\\
               -\sin\theta_3 & \cos\theta_3 & 0\\
                           0 &            0 & 1
              \end{pmatrix}.
\end{multline*}
Alternatively, one can write $M_1$ in a more succinct form by using
the adjoint action of the group, in terms of which we have $M_1 = \Ad
(e^{\theta_2 X_2} e^{-\theta_1 X_1}e^{\theta_3 X_3})$.  One can
therefore see that $M_1$ is indeed an element of $\SO(2,1)$, as we
obtained in Section~\ref{sec:boundary}, where the parameters are given
by the fields themselves.

\subsection{The $\SU(2)$ component}

We now turn to $S_2$, whose matrix of boundary conditions we denote by
$M_2$.  We use the parametrisation of $\SU(2)$ given by the second
expression in \eqref{eq:parg}, and compute the corresponding invariant
vielbeins $e_2$ and $\bar e_2$ which read
\begin{gather*}
e = \begin{pmatrix}
    \cos\phi_3  & \cos\phi_1\sin\phi_3 & 0\\
    -\sin\phi_3 & \cos\phi_1\cos\phi_3 & 0\\
                0 &            -\sin\phi_1 & 1
    \end{pmatrix}\\
\bar e = \begin{pmatrix}
         \cos\phi_2 & 0 & \cos\phi_1\sin\phi_2\\
                    0 & 1 & -\sin\phi_1\\
        -\sin\phi_2 & 0 & \cos\phi_1\cos\phi_2
         \end{pmatrix}.
\end{gather*}
From this we obtain the corresponding background metric,
$(G_2)_{\mu\nu} = {e^a}_{\mu} (\eta_2)_{ab}{e^b}_{\nu}$.  If we now
choose $\alpha_2=\sin\phi_1 d\phi_2 d\phi_3$ we find for the matrix of
the boundary conditions
\begin{multline*}
\small
M_2= \begin{pmatrix}
                 \cos\phi_2 & 0 & \sin\phi_2\\
                            0 & 1 & 0\\
                -\sin\phi_2 & 0 & \cos\phi_2
              \end{pmatrix}
\begin{pmatrix}
                1 &             0 & 0\\
                0 & \cos\phi_1  & \sin\phi_1\\
                0 & -\sin\phi_1 & \cos\phi_1 
              \end{pmatrix}\\
\times \small \begin{pmatrix}
                \cos\phi_3 & -\sin\phi_3 & 0\\
                \sin\phi_3 &  \cos\phi_3 & 0\\
                           0 &            0 & 1
              \end{pmatrix}.
\end{multline*}
Also here we can use the adjoint action of the group to write $M_2 =
\Ad (e^{\phi_2 Y_2} e^{-\phi_1 Y_1}e^{\phi_3 Y_3})$.  In this
case the matrix describing the boundary conditions is an element of
$\SO(3)$, as obtained in Section~\ref{sec:boundary}, and again the
parameters are given by the fields themselves.

By putting $M_1$ and $M_2$ together we obtain the matrix of classical
boundary conditions for the open string sigma model on $\AdS_3\times
S^3$.  It is important to remark that the classical boundary
conditions we obtained in this way are described by a field-dependent
automorphism of the corresponding Lie algebra, which preserves the
metric.  Thus, on the one hand, these configurations do preserve
conformal invariance.  On the other hand, they give rise to gluing
conditions which have a similar form with the type-N gluing conditions
introduced in Section~\ref{sec:boundary}, the only difference being
that, here, they are field-dependent.

We now finally come to the main point of this section, which is to
investigate the possibility of having D-branes in $\AdS_3\times S^3$
which fill the whole target space.  In order to see this, we must
analyse the eigenvalues of the matrix of boundary conditions.  We
therefore need to rewrite the Neumann boundary conditions
\eqref{eq:cBC} in a slightly different form:
\begin{equation*}
  \d X^{\mu}= \tilde {M^\mu}_{\nu}\bar\d X^{\nu}~,
\end{equation*}
where the field-dependent matrix $\tilde M$ is given by
\begin{equation*}
  \tilde M = \frac{G+ \alpha}{G -\alpha}~.
\end{equation*}
Clearly, in order to have a consistent configuration, $\tilde M$
should not possess $-1$ eigenvalues, which would correspond to
Dirichlet directions.  An explicit calculation shows that $\tilde M_1$
possesses no $-1$ eigenvalues, whereas $\tilde M_2$ does possess $-1$
eigenvalues, but only at the points where our parametrisation of
$\SU(2)$ is singular, that is when $\cos\phi_1=0$.  In order to
explain these different results, we must take into account that, in
spite of our similar treatment of the two groups $\SL(2,\R)$ and
$\SU(2)$, there is however one important distinction between them:
$\SU(2)$ is a compact group, whereas $\SL(2,\R)$ is noncompact.
Consequently the $\SU(2)$ part of our sigma model analysis involves
some subtleties, for instance our parametrisation of $\SU(2)$ in
\eqref{eq:parg} becomes singular for $\cos\phi_1=0$, and the two-form
$B$ cannot be not globally defined.  This indicates that, even if we
impose Neumann boundary conditions in all directions the corresponding
D-brane will not fill the whole $S^3$, but rather a three-dimensional
submanifold of $S^3$.

In fact, one can show that these singular points make up a disjoint
union of two circles inside $S^3$, since
\begin{equation*}
g_2(\pm\pi/2,\phi_2,\phi_3) = e^{\pm\pi/2 Y_1}
e^{(\phi_3\mp\phi_2)Y_3}~,
\end{equation*}
and hence the corresponding D3-brane in $S^3$ is given by the
complement of the above circles in $S^3$.  Although a direct
comparison is not easy, this agrees at least morally with \cite{KlS}.
Hence we must conclude that one can construct D-brane configurations
which fill the whole group manifold $\SL(2,\R)$, but not the $\SU(2)$
manifold.

\section{The $N{=}1$ supersymmetric extension}
\label{sec:susyext}

Let us start by introducing the $N{=}1$ supersymmetric extension of
the affine Lie algebra $\widehat\gg$, which we will denote by
\begin{equation*}
  \widehat\gg_{N{=}1} = \widehat{\sl}(2,\R)_{N{=}1} \oplus
  \widehat{\su}(2)_{N{=}1}~,
\end{equation*}
with generators $(I_a,\psi_a)$ for the $\widehat{\sl}(2,\R)_{N{=}1}$
piece satisfying
\begin{align}
I_a (z)I_b (w) &= \frac{k(\eta_1)_{ab}}{(z-w)^2} + \frac{{f_{ab}}^c
                  I_c (w)}{z-w} + \reg~,\label{eq:sl21}\\
I_a (z)\psi_b (w) &= \frac{{f_{ab}}^c \psi_c (w)}{z-w} +
\reg~,\label{eq:sl22}\\
\psi_a (z)\psi_b (w) &= \frac{k(\eta_1)_{ab}}{z-w} +
                        \reg~.\label{eq:sl23}
\end{align}
and $(J_a,\chi_a)$ for the $\widehat{\su}(2)_{N{=}1}$ piece
satisfying
\begin{align}
J_a (z)J_b (w) &= \frac{k(\eta_2)_{ab}}{(z-w)^2} +
         \frac{{f_{ab}}^c J_c (w)}{z-w} + \reg~,\label{eq:su21}\\
J_a (z)\chi_b (w) &= \frac{{f_{ab}}^c \chi_c (w)}{z-w} +
         \reg~,\label{eq:su22}\\
\chi_a (z)\chi_b (w) &= \frac{k(\eta_2)_{ab}}{z-w} +
         \reg~.\label{eq:su23}
\end{align}
Apart from this, we also have the contribution of the free fields
$(\varphi_i,\lambda_i)$ on $T^4$, with the standard OPEs
\begin{align}
\d\varphi_i(z)\d\varphi_j(w) &= \frac{\delta_{ij}}{(z-w)^2} +
\reg~,\label{eq:sT41}\\
\lambda_i(z)\lambda_j(w) &= \frac{\delta_{ij}}{z-w} +
\reg~.\label{eq:sT42}
\end{align}

Then the generators of the $N{=}1$ SCA will be given by
\begin{align*}
\sT(z) =& \tfrac{1}{2k}\eta_1^{ab}(\tilde I_a \tilde I_b) +          
          \tfrac{1}{2k}\eta_2^{ab}(\tilde J_a \tilde J_b) + 
          \half\sum_{i=1}^{4} (\d\varphi_i\d\varphi_i) +\\
        &+\tfrac{1}{2k} \eta_1^{ab} (\d\psi_a\psi_b) +
          \tfrac{1}{2k} \eta_2^{ab} (\d\chi_a\chi_b) +
          \half\sum_{i=1}^{4} (\d\lambda_i\lambda_i)~,\\ 
\sG(z) =& \tfrac{1}{k}\eta_1^{ab}(\tilde I_a\psi_b) +
          \tfrac{1}{k}\eta_2^{ab}(\tilde J_a\chi_b) + 
          \sum_{i=1}^{4} (\d\varphi_i\lambda_i)\\
        &-\tfrac{1}{6k^2} f^{abc} (\psi_a \psi_b \psi_c) - 
          \tfrac{1}{6k^2} f^{abc} (\chi_a \chi_b \chi_c)~,
\end{align*}
where
\begin{equation*}
  \tilde I_a \equiv I_a - \tfrac{1}{2k}\eta_1^{bd}{f_{ab}}^c
  (\psi_c\psi_d)\qquad\text{and}\qquad \tilde J_a \equiv J_a -
  \tfrac{1}{2k}\eta_2^{bd}{f_{ab}}^c (\chi_c\chi_d)~.
\end{equation*}
This SCFT has a central charge $c=15$.  Notice that although so far we
have only considered the holomorphic sector, we have a similar
structure for the antiholomorphic sector as well.  In other words, we
have a $(1,1)$ SCFT.

As in the bosonic case, we consider two classes of gluing conditions.
The gluing conditions of type-N are given by
\begin{equation}
  \label{eq:sNbc}
  J_a(z) + {R^b}_a {\bar J}_b(\bar z) = 0~,\qquad 
  \psi_a(z) + {S^b}_a {\bar\psi}_b(\bar z) = 0~,
\end{equation}
whereas the gluing conditions of type-D read
\begin{equation}
  \label{eq:sDbc} 
  J_a(z)- {R^b}_a {\bar J}_b(\bar z) = 0~,\qquad 
  \psi_a(z) - {S^b}_a {\bar\psi}_b(\bar z) = 0~.
\end{equation}
In both cases the coefficients ${R^b}_a$ and ${S^b}_a$ are defined by
$R,S:\gg\to\gg$, with $R(Z_a)=Z_b{R^b}_a$ and $S(Z_a)=Z_b{S^b}_a$, for
any $Z_a$ in $\gg$.  These conditions are to be understood as
supersymmetric generalisations of the gluing itions written down in
Section~\ref{sec:boundary}; therefore, $R$ is taken to be an
automorphism of $\gg$ which preserves the metric.  Moreover, since we
want to obtain supersymmetric configurations, the gluing conditions
satisfied by the fermions will undoubtedly be related to the ones of
the bosons; however we do not impose here any specific conditions on
$S$.

These gluing conditions have to satisfy a similar consistency
requirement as in the bosonic case.  In this context, consistency
means that the holomorphic SCFT is set equal to the antiholomorphic
SCFT up to an automorphism of the $N{=}1$ SCA; in other words, at the
boundary we must have
\begin{equation*}
  \label{eq:sci}
  \sT(z) = \bar\sT(\bar z)\qquad\text{and}\qquad \sG(z) =
  \pm\bar\sG(\bar z)~.
\end{equation*}
These boundary conditions have been written down previously in
\cite{SDKS}, in the context of Kazama--Suzuki models.

The first requirement translates into a number of conditions on the
matrices $R$ and $S$.  Let us start with the type-D boundary
conditions.  From the quadratic terms in the currents we obtain that 
\begin{equation*}
R^T \eta R = \eta~,\qquad S^T \eta R = \pm\eta~,
\end{equation*}
which immediately implies that
\begin{equation}
S = \pm R~,\label{eq:sr}
\end{equation}
as one would expect from supersymmetry.  Further, from the cubic terms 
in the currents we have that
\begin{equation*}
[S(Z_a),S(Z_b)] = \pm S([Z_a,Z_b])~,\qquad 
[R(Z_a),S(Z_b)] = S([Z_a,Z_b])~,
\end{equation*}
which, together with \eqref{eq:sr}, implies that
\begin{equation}
[R(Z_a),R(Z_b)] = R([Z_a,Z_b])~.
\end{equation}
In other words, the conditions that $R$ must satisfy in order for the
corresponding type-D configurations to preserve superconformal
invariance match exactly the assumptions already made on $R$.
Furthermore, it follows that these gluing conditions preserve the
infinite-dimensional symmetry of the $N{=}1$ current algebra
\eqref{eq:sl21}-\eqref{eq:su23}. 

If we turn now to the type-N gluing conditions, we obtain once again 
that $S=\pm R$, from the quadratic terms of $\sT$ and $\sG$.  However, 
from the cubic terms we obtain
\begin{equation*}
[S(Z_a),S(Z_b)] = \mp S([Z_a,Z_b])~,\qquad 
[R(Z_a),S(Z_b)] = - S([Z_a,Z_b])~,
\end{equation*}
which, together with \eqref{eq:sr}, implies that
\begin{equation}
[R(Z_a),R(Z_b)] = - R([Z_a,Z_b])~.
\end{equation}
This implies that $R$ must be an anti-automorphism, contrary to our
Ansatz.  From this we deduce that the type-N gluing conditions do
\emph{not} preserve the $N{=}1$ superconformal invariance of the
background.  For this reason, in the remaining of this paper, we will
concentrate on the type-D configurations.

From the bosonic case we know that $R$ takes a block-diagonal form
\eqref{eq:R}; therefore the condition \eqref{eq:sr} implies hat $S$
must have a similar form
\begin{equation}\label{eq:S}
S = \pm \begin{pmatrix}
            R_1 & 0\\
            0   & R_2
         \end{pmatrix}~.
\end{equation}
This shows that every bosonic type-D configurations that we determined
previously can be made into an $N{=}1$ supersymmetric configuration
without having to impose additional conditions.

\section{Spacetime supersymmetry}
\label{sec:spacetimesusy}

One of the most important properties of D-brane configurations
is that they preserve some of the spacetime supersymmetry of the
background in which they live, which translates into the fact that
they satisfy the BPS condition.  In the context of superconformal
field theories spacetime supersymmetry appears as a by-product of
$N{=}2$ superconformal invariance, being related, via bosonisation, to
the $U(1)$ current.  Instead of following this standard approach, here
we will analyse the spacetime symmetry preserved by the D-branes we
found using a different route, which was described in \cite{GKS}.

We will therefore consider the spacetime supercharges to be
constructed directly from the $N{=}1$ SCFT, by choosing five fermion
bilinears and bosonising them into five scalar fields $H_I$, with
$I=1,\ldots,5$ as follows
\begin{equation*}
\d H_1 = \frac{1}{k}(\psi_1\psi_2)~,\qquad 
\d H_2 = \frac{1}{k}(\chi_1\chi_2)~,\qquad 
i\d H_3 = \frac{1}{k}(\psi_3\chi_3)~,
\end{equation*}
\begin{equation*}
  \d H_4 = (\lambda_1\lambda_2)~,\qquad 
\d H_5 = (\lambda_3\lambda_4)~.
\end{equation*}
The corresponding spacetime supercharges will then read \cite{FMS}
\begin{equation}
Q = \oint dz e^{-\frac{\phi}{2}} S(z)~,
\end{equation}
where $\phi$ is the scalar field which appears in the bosonised
superghost system of the fermionic string, and
\begin{equation}
S(z) = e^{\frac{i}{2}\sum_I \epsilon_I H_I}~,
\end{equation}
is a linear combination of the spin fields, where the coefficients
$\epsilon_I = \pm 1$ label the possible supercharges, subject to a
number of requirements (for a detailed discussion see \cite{GKS}).
Thus, due to the requirement of mutual locality between the various
supercharges, and of BRST invariance these coefficients must satisfy
the following conditions:
\begin{equation*}
\epsilon_1 \epsilon_2 \epsilon_3 = 1~,
\qquad \epsilon_4 \epsilon_5 = 1~.
\end{equation*}
This yields eight supercharges for each of the holomorphic and
antiholomorphic sectors of the superstring background, which are
displayed in Table~\ref{tab:Q}.

\begin{table}[h!]
\centering
\renewcommand{\arraystretch}{1.1}
\begin{tabular}{|>{$}c<{$}|>{$}c<{$}|>{$}c<{$}|>{$}c<{$}|>{$}c<{$}|>{$}c<{$}|}
\hline
     & \epsilon_1 & \epsilon_2 & \epsilon_3 & \epsilon_4 &
       \epsilon_5\\ 
\hline
Q_1 & + & + & + & + & +\\
Q_2 & + & - & - & + & +\\
Q_3 & - & + & - & + & +\\
Q_4 & - & - & + & + & +\\
\hline
\end{tabular}
\qquad
\begin{tabular}{|>{$}c<{$}|>{$}c<{$}|>{$}c<{$}|>{$}c<{$}|>{$}c<{$}|>{$}c<{$}|}
\hline
     & \epsilon_1 & \epsilon_2 & \epsilon_3 & \epsilon_4 &
       \epsilon_5\\ 
\hline
Q_5 & + & + & + & - & -\\
Q_6 & + & - & - & - & -\\
Q_7 & - & + & - & - & -\\
Q_8 & - & - & + & - & -\\
\hline
\end{tabular}
\vspace{8pt}
\caption{The spacetime supercharges of the holomorphic sector.}
\label{tab:Q}
\end{table}

Given a certain D-brane configuration, we can use the gluing
conditions of the fermionic fields in order to derive the boundary
conditions satisfyed by the supercharges, and determine, in this way,
the fraction of spacetime supersymmetry preserved by that particular
boundary state.  To illustrate, let us consider the case of the
configurations described by a matrix $R$, with $R_1$ of the form
\eqref{eq:R1i}.  The corresponding conditions satisfied by the
fermions read
\begin{align*}
\psi_1 - \cos\alpha \bar\psi_1 - \sin\alpha \bar\psi_2 = 0~,\qquad &
\chi_1 - \cos\beta \bar\chi_1 - \sin\beta \bar\chi_2 = 0~,\\
\psi_2 + \sin\alpha \bar\psi_1 - \cos\alpha \bar\psi_2 = 0~,\qquad &
\chi_2 + \sin\beta \bar\chi_1 - \cos\beta \bar\chi_2 = 0~,\\
\psi_3 - \bar\psi_3 = 0~,\qquad & \chi_3 - \bar\chi_3 = 0~,
\end{align*}
where we have systematically ignored a $\pm$ sign coming from
\eqref{eq:S}, which does not affect the fermion bilinears in the
expression of $S(z)$.  Therefore we deduce that $H_I = \bar H_I$, for
$I=1,2,3$.  Since we are considering D-branes embedded in
$\AdS_3\times S^3$, the boundary conditions corresponding to the
fermions on the torus are always the same (that is, Dirichlet), and
will therefore give $H_I = \bar H_I$, for $I=4,5$. From this it
follows that
\begin{equation}\label{eq:BPS}
Q_{\alpha} = \bar Q_{\alpha}~,\qquad\qquad \alpha=1,\ldots,8
\end{equation}
This means that all the corresponding D-brane configurations discussed
in Section~\ref{sec:Dgeometry} preserve half of the spacetime
supersymmetry of the background.

In order to analyse the spacetime supersymmetry properties of the
D-brane configurations described by matrices $R$ with $R_1$ of the
form \eqref{eq:R1ii}, we need to adopt a slightly different choice for
the five fermion bilinears and, thus, for the corresponding scalar
fields $H_I$:
\begin{equation*}
i\d H_1 = \frac{1}{k}(\psi_1\psi_3)~,\qquad 
\d H_2 = \frac{1}{k}(\chi_1\chi_2)~,\qquad 
\d H_3 = \frac{1}{k}(\psi_2\chi_3)~,
\end{equation*}
\begin{equation*}
  \d H_4 = (\lambda_1\lambda_2)~,\qquad 
\d H_5 = (\lambda_3\lambda_4)~.
\end{equation*}
Similarly, we obtain that for the corresponding D-brane configurations
the supercharges satisfy the same conditions \eqref{eq:BPS}.  Due to
the nonlocal nature of the dependence of the spacetime supercharges on
the fermionic fields and to the particular form of the boundary
conditions satisfied by the fields in the third case of the discussion
in Section~\ref{sec:Dgeometry} (that is, where $R_1$ is of the form
\eqref{eq:R1iii}), it is rather difficult to determine the fraction of
spacetime supersymmetry preserved this type of configurations.

To summarise, we have obtained that all the D-brane configurations
which preserve the superconformal invariance of te background and have 
a geometrical description give rise to BPS states preserving half of
the spacetime supersymmetry.


Let us conclude this section with a remark.  Every boundary state that 
we identified and which gives rise to a D$p$-brane in $\AdS_3\times
S^3$ (as shown in Table 2), can also describe (with appropriate
boundary conditions in the `flat' directions) D$(p+2)$- and
D$(p+4)$-branes wrapped on $T^2$ and $T^4$, respectively.

\section{Discussion}
\label{eq:discussion}

In this paper we have studied, using the SCFT framework and the
boundary state formalism, the possible D-brane configurations which
can be consistently defined in an $\AdS_3\times S^3$ background
characterised by a purely NS-NS B field.  We have seen that at the
bosonic level one can define two classes of gluing conditions which
preserve conformal invariance, which we called type--N and type--D
conditions.  Type--D gluing conditions have the additional property
that they preserve the infinite-dimensional symmetry of the bulk
theory generated by the chiral currents.  

In order to determine the geometry of the corresponding D-brane
configurations we had to first obtain the boundary conditions encoded
in the algebraic gluing conditions.  Then, by analysing these boundary
conditions, we were able to show that type--D gluing conditions
describe D-branes whose worldvolume is given by shifted conjugacy
classes in the group manifold.  Furthermore, this type of
configurations admits an $N{=}1$ supersymmetric generalisation which
preserves not only the superconformal invariance of the corresponding
background, but also the underlying $N{=}1$ affine superalgebra of the
bosonic and fermionic currents.  By contrast, type--N gluing
conditions do not preserve the current algebra of the bulk theory.
They describe D-brane configurations which are slightly more difficult
to characterise geometrically, in the most general case.  We have
however seen that we obtain in particular D-branes whose worldvolume
is an open submanifold of dimension equal to the dimension of the
target manifold, and also some subgroups and cosets.  This type of
D-brane configurations does not seem to generalise to the
supersymmetric case, in the sense that it does not yield
superconformal configurations.

The two classes of bosonic D-brane configurations found here bear a
certain degree of similarity with the D-brane configurations obtained
in \cite{KlS} using a open string WZW model analysis.  There it was
moreover shown that the two distinct classes of D-brane configurations
are related by Poisson--Lie T--duality.  It would be interesting to
invest igate this possible relashionship also in our setting.


All D-brane configurations whose spacetime supersymmetry properties we
have been able to analyse have in common the fact that the complex
structure (implicitly defined through the choice of fermion bilinears)
on the ten-dimensional background gives rise, when restricted to the
tangent space of the worldvolume of a given D-brane, to a complex
structure on the corresponding submanifold of the target.  In other
words, these D-brane configurations correspond to pseudocomplex cycles
in the sense of \cite{SDKS}.

\section*{Acknowledgements}

It is a pleasure to thank JM~Figueroa-O'Farrill and AA~Tseytlin for
many useful discussions and for a critical reading of the manuscript.
This work was supported by a PPARC Postdoctoral Fellowship.

\appendix

\section{Conjugacy classes of $\SL(2,\R)$}
\label{sec:conj-classes}
\vspace{-5pt}
\begin{center}
(with JM Figueroa-O'Farrill)
\end{center}
\vspace{10pt}

In this appendix we determine the conjugacy classes of the noncompact
Lie group $\SL(2,\R)$.  This is probably a classic result, but we are
unaware of any reference.  We also analyse their causal structure
relative to the natural bi-invariant metric on the group.

\subsection{Jordan normal forms}

We will think of $\SL(2,\R)$ as the group of $2\times 2$ real
matrices with unit determinant:
\begin{equation*}
  \SL(2,\R) := \left\{ 
      \begin{pmatrix}
        a & b \\ c & d
      \end{pmatrix} \biggl|\, a d - b c = 1 \right\}~.
\end{equation*}
This shows that $\SL(2,\R)$ is a three-dimensional Lie group, which
can be represented as a hyperboloid in $\R^4$.

Let us embed $\SL(2,\R)$ in $\GL(2,\R)$, and in this way think of
every element in $\SL(2,\R)$ as the matrix of a linear transformation
in $\R^2$ relative to some basis.  Conjugation by $\GL(2,\R)$ will
then correspond to a change of basis in $\R^2$.  The
$\GL(2,\R)$-orbits in $\SL(2,\R)$ will then be labelled by, say,
normal forms of the linear transformations.  One such normal form is
the \emph{Jordan normal form}.  Although this usually is presented in
a way that requires a complex change of basis---that is, conjugation
in $\GL(2,\C)$---it is easy to restrict oneself to real changes of
basis.

According to the main theorem in the theory of Jordan normal forms,
any $2\times 2$ complex matrix is conjugate under $\GL(2,\C)$ to one
of the following normal forms:
\begin{equation*}
  \begin{pmatrix}
    \lambda_1 & 0 \\ 0 & \lambda_2
  \end{pmatrix}
  \qquad\text{or}\qquad
  \begin{pmatrix}
    \lambda & 1 \\ 0 & \lambda
  \end{pmatrix}~.
\end{equation*}
If we start with a matrix which actually belongs to $\SL(2,\R)$, the
normal form must have unit determinant and real trace, since the trace 
and determinant are invariant under conjugation.  That means that the
normal forms of matrices in $\SL(2,\R)$ are of the form
\begin{equation*}
  \begin{pmatrix}
    \lambda & 0 \\ 0 & 1/\lambda
  \end{pmatrix}
  \qquad\text{or}\qquad
  \begin{pmatrix}
    \pm 1 & 1 \\ 0 & \pm 1
  \end{pmatrix}~,
\end{equation*}
where $\lambda + 1/\lambda$ is real.

If the resulting normal form is real, then it is plain to see that the
matrices are actually conjugate in $\GL(2,\R) \subset \GL(2,\C)$.
To see this let $M$ and $M'$ be real $2\times 2$ matrices which are
conjugate under $\GL(2,\C)$.  This means that there exists some
matrix $S\in\GL(2,\C)$ such that
\begin{equation*}
  M S = S M'~.
\end{equation*}
We can interpret this as a system of homogeneous linear equations for
the entries of $S$ with real coefficients.  Because the constraint
that the determinant of $S$ be nonzero is an open condition, it means
that we can choose the entries of $S$ to be real.  (Of course, there
will be other choices of $S$ which are complex.)

If the resulting normal form is not real, then it is necessarily of
the form
\begin{equation*}
  \begin{pmatrix}
    e^{i\theta} & 0 \\ 0 & e^{-i\theta}
  \end{pmatrix}\qquad\text{with $\theta$ real.}
\end{equation*}
This matrix is itself conjugate under $\GL(2,\C)$ to the real matrix
\begin{equation*}
  \begin{pmatrix}
    \cos\theta & \sin\theta \\ -\sin\theta & \cos\theta
  \end{pmatrix}~.
\end{equation*}
Hence the original matrix in $\SL(2,\R)$ is conjugate to the above
matrix under $\GL(2,\C)$ and by the previous argument, the
conjugation can actually be taken to be in $\GL(2,\R)$.

In summary, the $\GL(2,\R)$ orbits of $\SL(2,\R)$ are labelled by
the following matrices
\begin{equation}
  \label{eq:gl-orbits}
  \begin{pmatrix}
    \lambda & 0 \\ 0 & 1/\lambda
  \end{pmatrix}\qquad
  \begin{pmatrix}
    \cos\theta & \sin\theta \\ -\sin\theta & \cos\theta
  \end{pmatrix}\qquad\text{or}\qquad
  \begin{pmatrix}
    \pm 1 & 1 \\ 0 & \pm 1
  \end{pmatrix}~,
\end{equation}
where we can choose $\theta$ in $[0,\pi]$ and $\lambda$ real with
$0<|\lambda|\leq 1$.  These choices correspond to the choice in ordering 
the eigenvalues of the (diagonalisable) normal forms.

In order to recover the $\SL(2,\R)$ conjugacy classes from these
$\GL(2,\R)$ orbits it is necessary to decompose every $\GL(2,\R)$
orbit into $\SL(2,\R)$ orbits.  Let $M$ be a matrix in $\SL(2,\R)$
and let $\eO$ denote its $\GL(2,\R)$ orbit:
\begin{equation*}
  \eO := \left\{ g M g^{-1} \mid g \in \GL(2,\R) \right\}~.
\end{equation*}
The Lie group $\GL(2,\R)$ has two connected components corresponding
to those elements with positive and negative determinant.
Correspondingly $\eO$ breaks up into two connected components: $\eO =
\eO_+ \cup \eO_-$, where
\begin{equation*}
  \eO_\pm := \left\{ g M g^{-1} \mid g \in \GL(2,\R)~\text{and}~\pm
  \det g > 0 \right\}~.
\end{equation*}
Now every matrix $g\in\GL(2,\R)$ with $\det g>0$ can be written as
\begin{equation*}
  g = \sqrt{\det g}\,s\quad\text{where}\quad s := \frac{1}{\sqrt{\det g}}\,g~,
\end{equation*}
where the matrix $s$ has unit determinant and hence belongs to
$\SL(2,\R)$.  Since for such a matrix $g$,
\begin{equation*}
  g M g^{-1} = s M s^{-1}~,
\end{equation*}
we see that $\eO_+$ is precisely the $\SL(2,\R)$-orbit of $M$.

On the other hand, let $g\in\GL(2,\R)$ with $\deg g <0$.  Then we can 
write it as
\begin{equation*}
  g = \sqrt{|\det g|}\,s\,t~,
\end{equation*}
where $t$ is any matrix with determinant $-1$, for example
\begin{equation*}
  t:=\begin{pmatrix} 1 & 0 \\ 0 & -1 \end{pmatrix}~,
\end{equation*}
and where $s$ is now given by
\begin{equation*}
 s := \frac{1}{\sqrt{|\det g|}}\,g\, t^{-1}~,
\end{equation*}
and has unit determinant again.  Now for such a $g$, we have that
\begin{equation*}
  g M g^{-1} = s t M t^{-1} s^{-1}~,
\end{equation*}
whence $\eO_-$ is the $\SL(2,\R)$-orbit of the matrix $t M t^{-1}$
which belongs to $\SL(2,\R)$.

In summary we see that the $\GL(2,\R)$-orbit of a matrix $M \in
\SL(2,\R)$ breaks up in at most two $\SL(2,\R)$ orbits: that of $M$
and that of $t M t^{-1}$.  It might happen that $M$ and $t M t^{-1}$
are actually in the same $\SL(2,\R)$ orbit, and one has to check this
case by case.  At any rate, it is now a simple matter to enumerate the
conjugacy classes of $\SL(2,\R)$ from the enumeration of the
$\GL(2,\R)$ orbits in \eqref{eq:gl-orbits}.  Every matrix in
$\SL(2,\R)$ is conjugate in $\SL(2,\R)$ to one of the following
matrices:
\begin{equation}
  \label{eq:sl-orbits}
  \begin{pmatrix}
    \lambda & 0 \\ 0 & 1/\lambda
  \end{pmatrix}\quad
  \begin{pmatrix}
    \cos\theta & \sin\theta \\ -\sin\theta & \cos\theta
  \end{pmatrix}\quad
  \pm\begin{pmatrix}
    1 & 1 \\ 0 & 1
  \end{pmatrix}\quad\text{or}\quad
  \pm\begin{pmatrix}
    1 & -1 \\ 0 & 1
  \end{pmatrix}~,
\end{equation}
where to avoid repetition we must now take $\lambda$ real with
$0<|\lambda|<1$ and $\theta\in [0,2\pi)$.

The results are summarised in Table~\ref{tab:conj-classes}.
There are two conjugacy classes $\eC_{\pm e}$ consisting of a point
each, corresponding to the elements $\pm e$ in the centre of
$\SL(2,\R)$.  The remaining conjugacy classes are two-dimensional:
four one-parameter families corresponding to $\eC_\lambda$ for
$\lambda\in(0,1)$ and $\lambda\in(-1,0)$ and to $\eC_\theta$ for
$\theta\in(0,\pi)$ and $\theta\in(-\pi,0)$, and four isolated classes
$\eC_{\pm\pm}$.

\begin{table}[h!]
\centering
\setlength{\extrarowheight}{3pt}
\begin{tabular}{|>{$}c<{$}|>{$}c<{$}|>{$}c<{$}|c|}\hline 
\text{Class} & \text{Element} & \text{Topology} & Causal type\\ \hline\hline
\eC_e & \begin{pmatrix} 1 & 0\\ 0 & 1 \end{pmatrix} & \text{point} &\\
\eC_{-e} & \begin{pmatrix} -1 & 0\\ 0 & -1 \end{pmatrix} & \text{point} &\\
\eC_{++} & \begin{pmatrix} 1 & 1\\ 0 & 1\end{pmatrix} & \R\times S^1 
& degenerate\\
\eC_{--} & \begin{pmatrix} -1 & -1\\ 0 & -1\end{pmatrix} & \R
\times S^1 & degenerate\\
\eC_{+-} & \begin{pmatrix} 1 & -1\\ 0 & 1\end{pmatrix} & \R \times
S^1 & degenerate\\
\eC_{-+} & \begin{pmatrix} -1 & 1\\ 0 & -1\end{pmatrix} & \R \times
S^1 & degenerate\\
\eC_\theta & \begin{pmatrix} \cos\theta & \sin\theta\\ -\sin\theta &
\cos\theta \end{pmatrix} & \R^2 & euclidean\\
\eC_\lambda & \begin{pmatrix} \lambda & 0\\ 0 & 1/\lambda \end{pmatrix} &
\R \times S^1 & minkowskian\\
\hline
\end{tabular}
\vspace{8pt}
\caption{$\SL(2,\R)$ conjugacy classes, with typical element, topology 
  and causal type.}
\label{tab:conj-classes}
\end{table}

\subsection{The geometry of the conjugacy classes}

\begin{figure}[h!]
\centering
\mbox{%
\subfigure[$|x|<1$]{\includegraphics[width=0.3\textwidth]%
{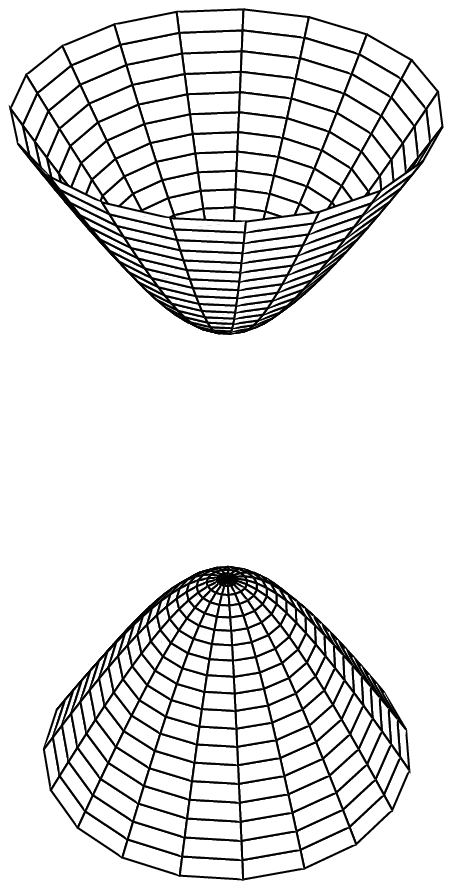}}
\subfigure[$|x|=1$]{\includegraphics[width=0.3\textwidth]%
{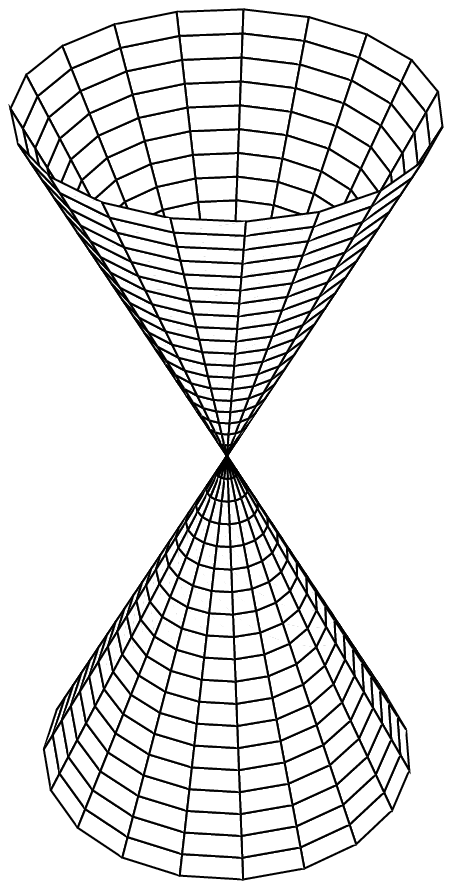}}
\subfigure[$|x|>1$]{\includegraphics[width=0.3\textwidth]%
{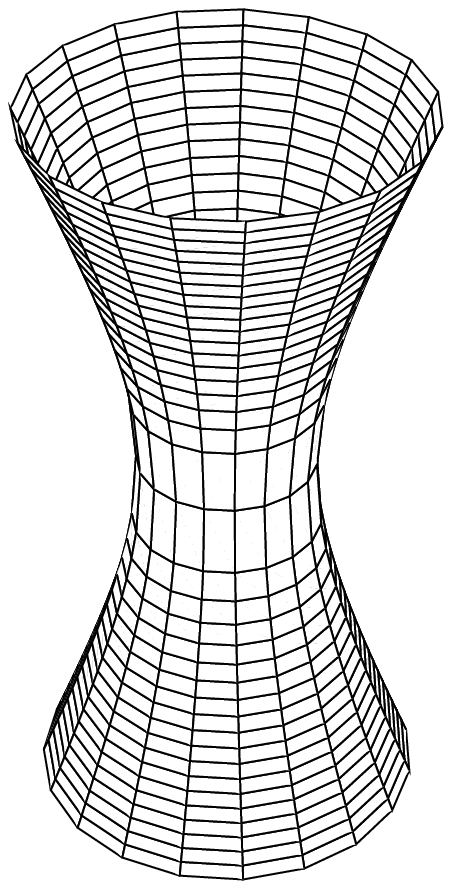}}
}
\caption{The different types of conjugacy classes of $\SL(2,\R)$.}
\label{fig:conj-classes}
\end{figure}

It is possible to understand the geometry of these conjugacy classes
in $\SL(2,\R)$.  To this end we let us reconsider the embedding of
$\SL(2,\R)$ as a hyperboloid in $\Mat(2,\R) \cong \R^4$, this time
using a different coordinate system for $\R^4$:
\begin{equation*}
  \SL(2,\R) = \left\{ \begin{pmatrix}
    x + u & y + v\\ y - v & x - u
  \end{pmatrix} \biggr|~ x^2 + y^2 = 1 + u^2 + v^2 \right\}~.
\end{equation*}
This embedding has the virtue of exhibiting the $\SO(2,2)$ isometry
group of $\SL(2,\R)$ manifestly.  The isometries are nothing but the
product of left and right transformations (modulo the centre which
acts trivially):
\begin{equation*}
  \SO(2,2) \cong \frac{\SL(2,\R) \times \SL(2,\R)}{\Z_2}~.
\end{equation*}
For our present purpose, a more interesting property of this embedding
is that the coordinate $x$ equals half the trace, which is an
invariant of the conjugacy class.  Therefore the conjugacy classes
will be contained in the intersection of the three-dimensional
hyperboloid
\begin{equation}
  \label{eq:H3}
H_3~: \qquad  x^2 + y^2 = 1 + u^2 + v^2
\end{equation}
with the affine hyperplanes $x=\text{constant}$.  We can distinguish
several regions of interest: $|x|<1$, $|x|=1$, and $|x|>1$.  We will
now analyse each of them in turn.  The results are illustrated in
Figure~\ref{fig:conj-classes}.

We start with the regions $|x|>1$.  The intersection of the affine
hyperplane $x=\text{constant}$ with $H_3$ is given by $(x^2-1) + y^2 =
u^2 + v^2$ which is a two-dimensional one-sheeted hyperboloid.  It is
connected and hence it is a conjugacy class: the class $\eC_\lambda$,
where $x = \lambda + 1/\lambda$.

The intersection of $H_3$ with the affine hyperplane $x=\pm 1$ is the
light-cone $y^2 = u^2 + v^2$.  This breaks up into three conjugacy
classes: the apex of the cone, which consists of the class
$\eC_{\pm e}$, the upper light-cone $y>0$ with the apex removed, which
is the class $\eC_{\pm +}$ and the lower light-cone $y<0$ with the apex 
removed, which is the class $\eC_{\pm -}$.

Finally, for $|x|<1$, the intersection of $H_3$ with the affine
hyperplane $x=\text{constant}$ is the two-dimensional two-sheeted
hyperboloid $y^2 = u^2 + v^2 + (1-x^2)$.  Each sheet is one conjugacy
class.  If we let $x = \cos\theta$ then the upper sheet is the class
$\eC_\theta$ when $\theta\in(0,\pi)$ and the lower sheet is the class
$\eC_\theta$ when $\theta\in(-\pi,0)$.

As a concluding comment, let us remark that the method presented here can
be employed with a little extra effort to enumerate the conjugacy
classes of $\SL(n,\R)$.

\subsection{The causal structure of the conjugacy classes}

Due to their possible interpretation as D-branes, it is important
to establish the causal structure of the conjugacy classes which were
found above: only those classes which are nondegenerate can be
straightforwardly interpreted as boundary conditions for strings.  The
determination of the causal structure is made easy by the fact that
these classes are described by the intersection of affine hyperplanes
in $\R^4$ with the hyperboloid $H_3$ in \eqref{eq:H3} defining the
embedding of $\SL(2,\R)$ in $\R^4$.  As was mentioned above, this
embedding is isometric provided we endow $\R^4$ with a split metric
of signature $(2,2)$.  In the coordinates $(x,y,u,v)$ chosen above,
such a metric is given by
\begin{equation}
  \label{eq:22metric}
  ds^2 = du^2 + dv^2 - dx^2 - dy^2~.
\end{equation}
It is then a simple matter to work out the induced metric on the
conjugacy classes.  Let us now summarise the results.  Of course, we
only need concern ourselves with those conjugacy classes which are not 
pointlike.

\subsubsection{$\eC_{\pm\pm}$}

These conjugacy classes are the deleted halves of the light-cones at
$x=\pm1$.  They are defined by this equation together with $y^2 = u^2
+ v^2$.  Let us parametrise the conjugacy class by $(\varrho,
  \vartheta)$ in the following way:
\begin{equation*}
  x = \pm 1\quad
  y = \pm \varrho\quad
  u = \varrho \cos\vartheta\quad
  v = \varrho \sin\vartheta~.
\end{equation*}
The induced metric is then given by
\begin{equation*}
  ds^2 = \varrho^2 d\vartheta^2~,
\end{equation*}
which is clearly degenerate.  This means that the conjugacy classes
$\eC_{\pm\pm}$ (with signs uncorrelated) cannot be interpreted as
D-branes, at least straightforwardly.

\subsubsection{$\eC_\theta$}

These are two-sheeted hyperboloids obtained by intersecting the affine 
hyperplane defined by constant $x$ with $|x|<1$ and the hyperboloid
$H_3$.  We parametrise these classes by $(\varrho, \vartheta)$ in
the following way
\begin{equation*}
  y = \pm \sqrt{\varrho^2 + (1-x^2)}\quad
  u = \varrho \cos\vartheta\quad
  v = \varrho \sin\vartheta~.
\end{equation*}
The induced metric is then given by
\begin{equation*}
  ds^2 = \varrho^2 d\vartheta^2 + \frac{(1-x^2)}{\varrho^2 +
  (1-x^2)}\, d\varrho^2~,
\end{equation*}
which is clearly euclidean.  Therefore the corresponding D-branes
are euclidean D-strings.

\subsubsection{$\eC_\lambda$}

These are the one-sheeted hyperboloids obtained by intersecting the affine 
hyperplane defined by constant $x$ with $|x|>1$ and the hyperboloid
$H_3$.  We parametrise these classes by $(y, \vartheta)$ in
the following way
\begin{equation*}
  u = \sqrt{y^2 + (x^2-1)} \cos\vartheta\quad
  v = \sqrt{y^2 + (x^2-1)} \sin\vartheta~.
\end{equation*}
The induced metric is then given by
\begin{equation*}
  ds^2 = \left(y^2 + (x^2 -1)\right) d\vartheta^2 - \frac{(x^2
  -1)}{y^2 + (x^2 -1)}\, dy^2~,
\end{equation*}
which is clearly minkowskian.  Therefore the corresponding D-branes
are D-strings.

\providecommand{\href}[2]{#2}\begingroup\raggedright\endgroup
\end{document}